\documentclass[letterpaper, 10 pt, conference]{ieeeconf}  % Comment this line out if you need a4paper
\IEEEoverridecommandlockouts                              % This command is only needed if
\usepackage{epsfig} % for postscript graphics files
\usepackage{graphicx}
\usepackage{amsmath}
\usepackage{array}
\usepackage{amssymb}
\usepackage{verbatim}
\usepackage{cleveref}
\usepackage{color}
\usepackage{relsize}
\usepackage{tikz}
\usepackage{fancyhdr}
\usepackage{graphics}
\usepackage{setspace}
\usepackage{epstopdf}
\usepackage{multirow}
\usepackage{lipsum}
\usepackage{textcomp}
\usepackage{mathtools}
\usepackage{cite}
\usepackage{setspace}
\usepackage{booktabs}
\usepackage{breqn}
\usepackage{algorithm}
\usepackage[noend]{algpseudocode}
\usepackage{flushend}
\usepackage{multicol}
\usepackage{float}

\fancypagestyle{firstpage}{%
  \lhead{\vspace{-0.75 cm} \small \textbf{57th IEEE Conference on Decision and Control (CDC)\\December 17-19, 2018, Miami Beach, FL, USA}}
}

\title{\Large \bf
Two-Layer Model Predictive Battery Thermal and Energy Management Optimization for Connected and Automated Electric Vehicles$^{*}$}

\author{Mohammad Reza Amini$^{1}$, Jing Sun$^{1}$, and Ilya Kolmanovsky$^{2}$% <-this % stops a space
\thanks{*This paper is based upon the work supported by the United States Department of Energy (DOE), ARPA-E NEXTCAR program under award No. DE-AR0000797.}% <-this % stops a space
\thanks{$^{1}$Mohammad Reza Amini and Jing Sun are with the Department of Naval Architecture \& Marine Engineering, University of Michigan, Ann Arbor, MI 48109 USA. Emails: {\tt\small \{mamini,jingsun\}@umich.edu}}%
\thanks{$^{2}$Ilya Kolmanovsky is with  the Department of Aerospace Engineering, University of Michigan, Ann Arbor, MI 48109 USA. Email: {\tt\small ilya@umich.edu}}%
}

\renewcommand{\baselinestretch}{1}
\begin{document}

\maketitle
\thispagestyle{firstpage}
%\pagestyle{empty}
%\pagestyle{fancy}
%\fancyhead[C]{\fontsize{14}{12} \selectfont \textit{Submitted to the 2017 American Control Conference, preprint.}}
%\chead{Submitted to the 2017 American Control Conference, preprint.}

%\doublespacing
%%%%%%%%%%%%%%%%%%%%%%%%%%%%%%%%%%%%%%%%%%%%%%%%%%%%%%%%%%%%%%%%%%%%%%%%%%%%%%%%
\begin{abstract}
Future vehicles are expected to be able to exploit increasingly the connected driving environment for efficient, comfortable, and safe driving. Given relatively slow dynamics associated with the state of charge and temperature response in electrified vehicles with large batteries, a long prediction/planning horizon is needed to achieve improved energy efficiency benefits. In this paper, we develop a two-layer Model Predictive Control (MPC) strategy for battery thermal and energy management of electric vehicle (EV), aiming at improving fuel economy through real-time prediction and optimization. In the first layer, the long-term traffic flow information and an approximate model reflective of the relatively slow battery temperature dynamics are leveraged to minimize energy consumption required for battery cooling while maintaining the battery temperature within the desired operating range. In the second layer, the scheduled battery thermal and state of charge ($SOC$) trajectories planned to achieve long-term battery energy-optimal thermal behavior are used as the reference over a short horizon to regulate the battery temperature. Additionally, an intelligent online constraint handling (IOCH) algorithm is developed to compensate for the mismatch between the actual and predicted driving conditions and reduce the chance for battery temperature constraint violation. The simulation results show that, depending on the driving cycle, the proposed two-layer MPC is able to save $2.8-7.9\%$ of the battery energy compared to the traditional rule-based controller in connected and automated vehicle (CAV) operation scenario. Moreover, as compared to a single layer MPC with a long horizon, the two-layer structure of the proposed MPC solution reduces significantly the computing effort without compromising the performance. 
%, compared to a similar single-layer MPC with a long horizon, for real-time implementation.
% Predictive traffic information brought in by the connected driving environment provides opportunities for fuel saving and other benefits. Given the slow dynamics of the battery thermal management system of the electrified vehicles, however, those benefits cannot be capitalized unless a long prediction horizon is incorporated. In this paper, we present a two-layer Model Predictive Control (MPC) strategy for electric vehicles (EVs) battery thermal and energy management in a connected and automated vehicles (CAVs) environment, aiming to claim the fuel economy benefits through real-time prediction and optimization. In the first layer, the long-term traffic flow information is leveraged in exploring the relatively slow thermal dynamics to minimize energy consumption required for battery cooling while maintaining the temperature within the desired operating range. In the second layer, the scheduled battery thermal trajectories planned to achieve long-term battery optimal thermal behavior are used in a short horizon to regulate the battery temperature. The simulation results show that, compared to the traditional rule-based controller, the proposed two-layer MPC is able to save $2.8-7.9\%$ of the battery energy, depending on the driving cycle. Moreover, the two-layer structure of the proposed MPC reduces the significant computation efforts without compromising the performance, compared to a similar single-layer MPC with a long horizon, for real-time implementation.
\end{abstract}

%%%%%%%%%%%%%%%%%%%%%%%%%%%%%%%%%%%%%%%%%%%%%%%%%%%%%%%%%%%%%%%%%%%%%%%%%%%%%%%%
\vspace{-0.2cm}
\section{INTRODUCTION}\vspace{-0.1cm}
Efficient thermal management in electrified vehicles, including pure electric vehicles (EVs), hybrid electric vehicles (HEVs), and plug-in HEVs (PHEVs) is a significant factor in the overall vehicle energy consumption optimization. For EVs, cooling the battery pack in hot summer days takes substantial amount of energy, which can significantly compromise energy efficiency and driving range of pure EVs~\cite{zolot2002thermal}. Moreover, since the required power for cooling the battery pack is delivered by the battery itself, the operation of the battery thermal management (BTM) system directly interacts with other power loads, such as the traction power, forming intricate feedback loops. Therefore, optimizing the battery pack operating temperature during hot weather is essential for improving the overall vehicle energy efficiency. 

There are few works in the open literature dedicated to optimization of the EVs' BTM system, see~\cite{kim2014real} for an overview. The Pontryagin's maximum principle is used in~\cite{bauer2014thermal} to optimize the battery thermal and energy management dynamics. In a similar study~\cite{masoudi2015battery}, the dynamic programming (DP) approach has been used to find the global optimal solution for the BTM optimization problem. Both~\cite{bauer2014thermal} and \cite{masoudi2015battery} have shown the benefits of using optimization in minimizing the required cooling power to maintain the battery temperature within the optimal operating range. However, such an optimization is usually carried out offline under the assumption that the whole driving cycle is known a priori. 

While emerging connectivity and autonomous driving technologies are expected to provide unprecedented opportunities to improve mobility and safety, they also open up new dimensions for vehicle and powertrain control and optimization. Extensive studies have been carried out on fuel economy optimization for electrified vehicles~\cite{guanetti2018control}. However, the implications of the connected and automated vehicles (CAVs) operation on power and thermal management have not been fully explored for electrified vehicles. The CAV technology will allow for the incorporation of a range of new high-value information into the optimization process when determining the optimal energy flow and power split strategies under real-world driving conditions, thereby realizing the full (and hitherto unfulfilled) energy saving potential of electrified vehicles~\cite{lim2017distance}. %as well as the opportunities for the coordination between the on-board thermal management system and the powertrain control

%With respect to the potential applications of predictive control, 
The thermal subsystem of an EV has several special characteristics and requirements that are relevant to predictive control. In particular, due to slow thermal dynamics, the optimization has to consider a long time horizon.%To allow for an integrated and fully optimized operation of the vehicle system that accounts for the different time scales in the physical responses of the powertrain and battery pack, multi-scale optimization and prediction can be exploited.%Coordinating the short time scale and long time scale controls will require improved data communications and efficient control optimization. 
~{Inspired by the recent works in the literature on multi-layer optimization and prediction for systems with different time scales, including those for microgrids~\cite{cominesi2018two} and building energy management systems~\cite{lefort2013hierarchical}, we propose and develop an innovative two-layer MPC formulation for battery thermal and energy management optimization in this paper.} The proposed two-layer MPC cools the battery pack by keeping its temperature within a certain interval based on traffic conditions and vehicle power demand. Responding to the predicted traffic conditions provided by the CAV (V2V/V2I) environment, the proposed two-layer MPC manages the power used for cooling so that the powertrain system and associated subsystems operate efficiently to achieve overall system efficiency optimization for different driving scenarios. %\vspace{-0.2cm}

{The contribution of this paper is threefold. First, the energy saving potential of predictive BTM system in EVs is exploited by utilizing the traffic flow information over a long prediction horizon to account for relatively slow thermal dynamics of the battery. Second, a two-layer MPC framework is developed to not only reduce the computation complexity versus a single-layer MPC with a long horizon, but also to integrate the optimization of battery thermal and energy management into a hierarchical control framework to account for different time-scales of prediction and control at each layer. Third, an intelligent online constraint handling algorithm is developed and incorporated into the two-layer MPC structure to reduce the chances of the battery temperature constraint violation in the presence of the mismatch between the actual and predicted speed profiles, specifically when long range prediction is used.} \vspace{-0.1cm}

\vspace{-0.1cm}
\section{Battery Thermal and Electrical Models} \label{sec:1}
%\vspace{-0.1cm}
%
To formulate an MPC for BTM system, both the electrical and thermal characteristics of the battery have to be captured by the prediction model. The thermal sub-model approximates the battery pack as a lumped mass ($m_{bat}$) with heat capacity $C_{th, bat}$, and can be described as follows~\cite{bauer2014thermal}:
\vspace{-0.15cm}
\begin{gather}
\label{eqn:Eq1}
\dot{T}_{bat}(t)=\frac{1}{m_{bat} C_{th,bat}}(I_{bat}^2R_{bat}+\dot{Q}),
\end{gather}
where, $T_{bat}$, $I_{bat}$, and $R_{bat}$ are the battery temperature, current, and internal resistance, respectively. $\dot{Q}<0$ is the required heat flow rate for cooling the battery, and it is treated as the input to the battery thermal model in this paper. %In practice, $\dot{Q}$ is provided through a combination of air and liquid cooling loops in electric vehicles. The liquid cooling consist of an electric A/C compressor and a coolant loop. The air cooling regulates the battery temperature by blowing the compartment air into the battery pack using a fixed or variable speed fan. Since the compartment air is cooled by the A/C compressor, the battery air cooling loop is also coupled to the vehicle HVAC system.

The electric system sub-model of the battery includes the battery voltage ($U_{bat}$), which can be expressed as a function of the open-circuit voltage ($U_{oc}$), $R_{bat}$, and $I_{bat}$ as follows:  \vspace{-0.2cm}
\begin{gather}
\label{eqn:Eq2}
U_{bat}=U_{oc}-I_{bat}R_{bat}
\end{gather}
$U_{oc}$ and $R_{bat}$ in (\ref{eqn:Eq2}) are functions of the battery state-of-charge ($SOC$) and $T_{bat}$. The battery current can be written as a function of the total demanded power as follows:  \vspace{-0.2cm}
\begin{gather}
\label{eqn:Eq3}
I_{bat}=({P_{trac}+P_{temp}})/{U_{bat}},
\end{gather}
where, $P_{trac}$ is the demanded traction power, and $P_{temp}$ is the required power to provide $\dot{Q}$ for the BTM system. It is assumed in Eq.~(\ref{eqn:Eq3}) that $P_{trac}$ and $P_{temp}$ are the main power loads on the battery, and other auxiliary loads on the battery are neglected. By substituting $U_{bat}$ in~(\ref{eqn:Eq2}) with $(P_{trac}+P_{temp})/I_{bat}$ (Eq.~(\ref{eqn:Eq3})), $I_{bat}$ can be re-written: \vspace{-0.35cm}

\small
\begin{gather}
\label{eqn:Eq4}
I_{bat}(t)=\frac{U_{oc}-\sqrt{U_{oc}^2-4R_{bat}(P_{trac}+P_{temp})}}{2R_{bat}}
\end{gather}
\normalsize

Using (\ref{eqn:Eq4}), Eq.~(\ref{eqn:Eq1}) becomes: \vspace{-0.15cm}
\begin{gather}
\label{eqn:Eq6}
\dot{T}_{bat}(t)=\xi(T_{bat}(t))=\frac{\frac{(U_{oc}-\sqrt{U_{oc}^2-4R_{bat}P_{trac}+P_{temp}})^2}{4R_{bat}}+\dot{Q}}{m_{bat}C_{th,bat}}.
\end{gather}
Moreover, $SOC$ is governed by: \vspace{-0.2cm}
\begin{gather}
\label{eqn:Eq5}
\dot{SOC}(t)=\zeta({SOC}(t))=-\frac{I_{bat}(t)}{C_{nom}}
\end{gather}
where, $C_{nom}$ is the nominal capacity of the battery. %The demanded traction power can be estimated as~\cite{bauer2014thermal}:
%
% \begin{gather}
% \label{eqn:Eq7}
% P_{trac}=\begin{cases}
% \frac{V_{veh}(F_r+F_a+m\dot{V}_{veh}^2)}{\eta},~~~propelling\\
% {V_{veh}(F_r+F_a+m\dot{V}_{veh}^2)}{\eta},~~~braking
% \end{cases}
% \end{gather}
% %
% where $V_{veh}$ and $m$ are the vehicle speed and mass. $\eta=\eta(\tau,\omega)$ is the traction system efficiency, which is a function of the torque ($\tau$) and the rotation speed ($\omega$) of the electric motor, and it is calculated by using a look-up table. $F_r$  and $F_a$ are rolling and aerodynamic resistances calculated as follows: \vspace{-0.5cm}
% %
% \begin{gather}
% \label{eqn:Eq7_2}
% F_r=C_rmg,\\
% F_a=0.5\rho A_{f}C_dV_{veh}^2,
% \end{gather}
% %
%with $C_r$ and $C_d$ being the rolling resistance and aerodynamic drag coefficients, $A_f$ the vehicle frontal area, and $\rho$ the air density. In this paper, the road is assumed to be plain, i.e., the road grade is zero. 
$P_{temp}$ is modeled as a linear function of the heat flow rate ${\dot{Q}}$~\cite{bauer2014thermal}: \vspace{-0.2cm}
\begin{gather}
\label{eqn:Eq8}
P_{temp}(\dot{Q})=a_c\dot{Q},~~\dot{Q}\leq 0,~~a_c<0
\end{gather}
%along with $R_{bat}(SOC,T_{bat})$, $U_{oc}(SOC,T_{bat})$, and $\eta(\omega,\tau)$ look-up tables
where $a_c$ is a constant. The parameters of the battery and vehicle dynamics are adopted from the Autonomie~\cite{halbach2010model} software library for an electric vehicle.% These parameters are listed in Appendix~\ref{FirstAppendix}. %Moreover, the high-fidelity thermal and electrical models of the battery are adopted from Autonomie, and will be used as the virtual testbed to carry out the simulations. 
\vspace{-0.2cm}

\section{Single-Layer MPC for Battery Thermal and Power Management}\label{sec:3} \vspace{-0.1cm}
The traditional BTM system attempts to maintain the battery temperature at a specified level, without considering the traffic information. A setpoint within the optimal temperature range of the battery operation is selected and tracked. Since the battery temperature can increase rapidly due to aggressive vehicle acceleration and deceleration and BTM has limited bandwidth to respond, the desired battery temperature setpoint ($T_{bat}^{s.p.}$) is usually selected well below the upper limit of the optimum operation range to assure that the temperature stays within the limit during transients. This conservative tracking approach designed for worst case scenarios without predictive traffic information, often leads to considerable extra energy consumed for battery cooling. To exploit the traffic information made available through CAVs, and to improve thermal efficiency, a model predictive controller is designed in this section to minimize the required battery cooling power, and maintain the battery temperature within the desired operation range. 

\subsubsection{\textbf{Problem Formulation}}In order to formulate the MPC, first the thermal ($\xi$) and electric ($\zeta$) models (Eqs.~(\ref{eqn:Eq5}) and (\ref{eqn:Eq6})) are discretized by applying the Euler forward method: \vspace{-0.5cm}
% f_{SOC}(k) f_{T_{bat}}(k)

\small
\begin{gather}
\label{eqn:Eq9}
SOC(k+1)=f_{SOC}(k)=SOC(k)+T_s\zeta(SOC(k)),
%T_s\Big(\frac{U_{oc}-\sqrt{U_{oc}^2-4R_{bat}(P_{trac}+P_{temp})}}{2C_{nom}R_{bat}}\Big), \nonumber
\end{gather}
\vspace{-0.75cm}
\begin{gather}
\label{eqn:Eq10}
{T}_{bat}(k+1)=f_{T_{bat}}(k)={T}_{bat}(k)+T_s\xi(T_{bat}(k)),
%T_s\Big(\frac{\frac{(U_{oc}-\sqrt{U_{oc}^2-4R_{bat}P_{trac}+P_{temp}})^2}{4R_{bat}}+\dot{Q}}{C_{th,bat}}\Big), \nonumber
\end{gather}
\normalsize
where, $T_s$ is the sampling time for control update (e.g., $T_s$=1~sec). We consider a single-layer MPC with an economic cost function formulated over a finite-time horizon ($N$) with $\dot{Q}$ being the optimization variable: \vspace{-0.15cm}
%-------------------------------------
\begin{gather} \label{eqn:Eq11}
\begin{aligned}
& \underset{\dot{Q}(\cdot|k)}{\text{min}} & & \sum_{i=0}^{N} P_{temp}(i|k), \\
% \end{aligned}
% \end{gather}
% %
% \begin{gather} \label{eqn:Eq11}
% \begin{aligned}
%\tilde{\boldsymbol{y}}(k) & \\
& \text{s.t.}
& & T_{bat}(i+1|k)=f_{T_{bat}}(i|k),~{i=0,\cdots,N},\\
& 
& & SOC(i+1|k)=f_{SOC}(i|k),~{i=0,\cdots,N},\\
& 
& &T_{bat}^{LL}\leq T_{bat}(i|k)\leq T_{bat}^{UL},~{i=0,\cdots,N},\\
& 
& &30\% \leq SOC(i|k)\leq 90\%,~{i=0,\cdots,N},\\
& 
& &-3000~W\leq \dot{Q}(i|k)\leq 0,~{i=0,\cdots,N-1},\\
& 
& & T_{bat}(0|k)=T_{bat}(k),~SOC(0|k)=SOC(k),%\\
%& 
%& & \dot{Q}(0|k)=\dot{Q}(k)
\end{aligned}
\end{gather}
%-------------------------------------
%
{where, $(i|k)$ designates the prediction for the time instant $k+iT_s$ made at the time instant $k$.} The nonlinear MPC optimization problem (\ref{eqn:Eq11}) attempts to minimize the power spent for battery thermal managment $P_{temp}=a_c\dot{Q}$ over the prediction horizon, while enforcing the state and input constraints. $T_{bat}^{UL}$ and $T_{bat}^{LL}$ are the upper and lower limits of the battery operating temperature, and they are set to $40^oC$ and $20^oC$, respectively. Note that $\dot{Q}$ is always negative for battery cooling scenario. $f_{SOC}(k)$ and $f_{T_{bat}}(k)$ are the discritized nonlinear dynamics of $SOC$ and $T_{bat}$ calculated according to (\ref{eqn:Eq9}) and (\ref{eqn:Eq10}). The optimization problem is solved at every time step, then $\dot{Q}(k)$ is commanded to the system and the horizon is shifted by one step ($T_s$). The MPC simulation is carried out on a desktop computer, with an Intel\textsuperscript{\textregistered} Core i7@2.60 GHz processor, in MATLAB\textsuperscript{\textregistered}/SIMULINK\textsuperscript{\textregistered} using YALMIP~\cite{lofberg2004yalmip} for formulating the optimization problem, and IPOPT\cite{wachter2006implementation} for solving the optimization problem numerically. 

\subsubsection{\textbf{Performance Evaluation}}
\begin{figure}[h!]
	\begin{center}
		\includegraphics[width=7.35cm]%{Results/UDDS_T_bat_SingleLayer_MPC_Modified_Final_V2_1.eps}  \vspace{-0.34cm} 
{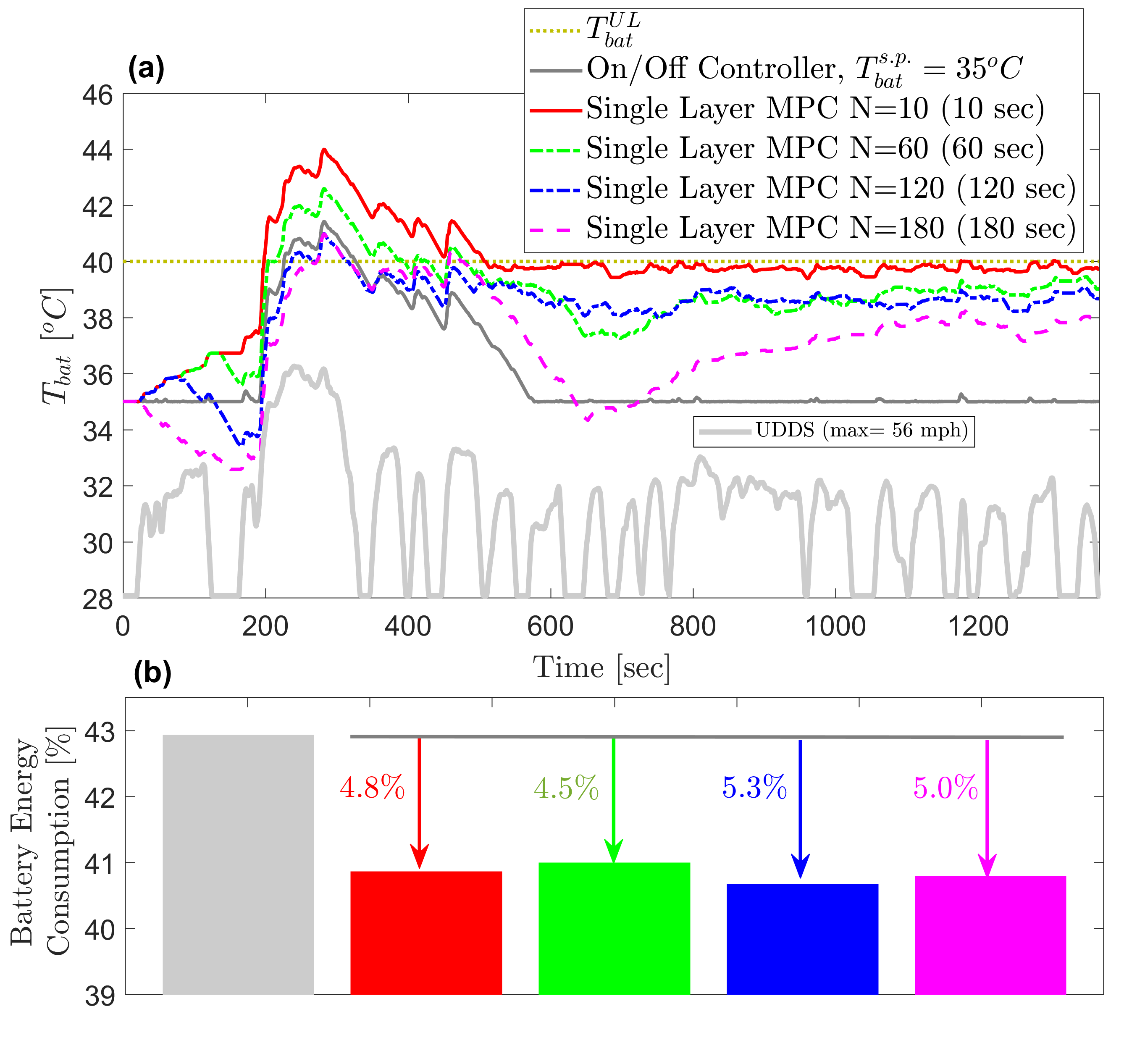}  \vspace{-0.72cm} 
\caption{The performance of the single layer MPC for battery thermal (a) and energy (b) management with different prediction horizons at $T_{bat,0}=35^oC$ for UDDS.}\vspace{-0.55cm} 
		\label{fig:UDDS_T_bat_SingleLayer_MPC_LongH_resize} 
	\end{center}
\end{figure}

Intuitively, the solution of the optimization problem in (\ref{eqn:Eq11}) results in the battery temperature to be close to the upper limit $T_{bat}^{UL}$ to minimize the cooling power consumption. However, since sudden changes in the $P_{trac}$ could increase the battery temperature and cause $T_{bat}$ to exceed the desired range, MPC with different prediction horizons will respond differently. Fig.~\ref{fig:UDDS_T_bat_SingleLayer_MPC_LongH_resize} shows the results of applying the nonlinear MPC (\ref{eqn:Eq11}) with different prediction horizons ($N$) on the EPA Urban Dynamometer Driving Schedule (UDDS), where it is assumed that the entire driving cycle is available, and the battery initial temperature ($T_{bat,0}$) is the same in all the simulated runs. % for all scenarios, and $T_{bat}^{UL}$ is fixed, and set to be $40^oC$~\cite{zolot2002thermal}. 
Additionally, the MPC simulation results are compared with simple rule-based (On/Off) controller which attempts to maintain $T_{bat}$ at a constant level of $T_{bat}^{s.p.}$=$35^oC$. %This setpoint is selected according to the setpoint given by the Autonomie software's BTM system~\cite{halbach2010model}.  
\vspace{-0.42cm}

Fig.~\ref{fig:UDDS_T_bat_SingleLayer_MPC_LongH_resize} shows that with a short prediction horizon ($N$=10), the controller does not have enough lead time to take mitigating actions to prevent constraint violation caused by sudden increase in the traction power around $t=200~sec$. However, as the prediction horizon is being extended, the MPC takes proactive actions to reduce the temperature before the heating load increases around $200~sec$, therefore significantly reducing the time for the battery to stay in over-temperature condition.~
%when the prediction horizon is $N=10$, the major rise in the demanded traction power around $t=200~sec$ becomes detectable to the MPC late. Thus, during the early seconds, the MPC minimizes the tempering power which leads to the increase in $T_{bat}$. Once the major rise in the traction power occurs, the battery temperature increases sharply, which leads the battery temperature hits $44^oC$. Eventually, when the battery is cooled down to below $T_{bat}^{UL}$, it will remain at $40^oC$ until the end of the driving cycle. Table~\ref{tab:Energy_Comparison} summarizes the results of single-layer MPC, in terms of energy saving and battery constraint violation, with the rule-based controller performance as the baseline. 
As shown in Fig.~\ref{fig:UDDS_T_bat_SingleLayer_MPC_LongH_resize}-b, compared to the simple rule-based controller, $4.5-5.3\%$ improvement in the energy consumption can be achieved by using the predictive controllers. The temperature upper limit violation, as expected, becomes less frequent as the prediction horizon being extended. When the horizon is longer than 120 $sec$, the MPC puts the efforts to decrease $T_{bat}$ from the early seconds, which eventually results in both fuel saving ($\geq5\%$) and reduced constraint violation by up to 41\% (compared to the rule-based controller). %\vspace{-0.25cm}% violation. up to 41\% reduction in $T_{bat}^{UL}$ violation, while up to 5\% saving in the energy consumption can be achieved. 

%It should be mention that an add-on logic has been applied to the single-Layer MPC to avoid infeasibility when the temperature constraint is violated. 
%\linespread{1.3}
% \begin{table}[h!]
% \begin{center}
% \small
% \centering
% \caption{Energy saving and battery constraint violation comparison of the Single-Layer MPC with different $N$ for Fig.~\ref{fig:UDDS_T_bat_SingleLayer_MPC_LongH_resize}. 
% \label{tab:Energy_Comparison}} %\vspace{0.4cm}
% \begin{tabular}{lcc}
%         \hline\hline
% ~~~~~~~~~~$controller$& $SOC$ (\%) & $T_{bat}^{UL}$ Violation (sec)  \\ \hline
% Rule-Based      & -~\textcolor{blue}{(Baseline)} & 101~\textcolor{blue}{(Baseline)}  \\ \hline
% MPC~$N=10$ & \textcolor{blue}{$+$4.8\%} & 321~\textcolor{red}{($\uparrow$218\%)} \\\hline
% %
% MPC~$N=60$ & \textcolor{blue}{$+$4.5\%} & 225~\textcolor{red}{$(\uparrow$123\%)} \\
% \hline
% %
% MPC~$N=120$ & \textcolor{blue}{$+$5.3\%} & 78~\textcolor{black!30!green}{($\downarrow$23\%)} \\
% \hline
% %
% MPC~$N=180$ & \textcolor{blue}{$+$5.0\%} & 60~\textcolor{black!30!green}{($\downarrow$41\%)} \\
% %
% % $~~~max~CPU$      &  &  &  &  \\
% % $time/iteration~[sec]$ & 2.40 & 4.51 & 9.57 & 18.47 \\
%         \hline\hline
%     \end{tabular}
% \end{center} \vspace{-0.6cm} 
% \end{table}
% \linespread{1}

\subsubsection{\textbf{Robustness to Prediction Uncertainty}}This study confirms the advantages of incorporating the future traffic information for battery thermal and energy optimization via an MPC framework. Due to the large thermal inertia and therefore large time constant in the thermal response, however, BTM system requires much longer horizon information to capitalize on the benefits associated with the MPC approach. %It was concluded that availability of information of future events helps to maintain the temperature within the desired range, and enhance the final state-of-charge of the battery. 
%While the results of the battery energy management ($SOC$) are almost the same for different prediction horizons, as $N$ becomes longer, the battery thermal behavior significantly changes, and the violation of the battery temperature operation limit reduces. This observation can be explained with respect to the difference in the time constants of the battery thermal and electrical systems. Compared to power management systems of the EVs, which deal with a relatively fast dynamic, and is usually performs in a short control horizon, it can be seen that the BTM system requires much longer horizon information. %This is because of the thermal inertia of the BTM system which makes the thermal dynamic relatively slow. Thus, the BTM systems require a much longer time horizon.  

On the other hand, the implementation of the MPC over a long horizon is not practical for two main reasons: (i) it is computationally demanding, and (ii) the accuracy in future traffic event prediction over a longer horizon cannot be guaranteed. The average %and maximum 
computation time per iteration and execution for the {single-layer} MPC with $N$=10,~60,~120,~180~$sec$ was recored as 0.75,~2.22,~6.24,~10.78~$sec$, respectively. It can be seen that the MPC with a long horizon is computationally expensive with average computation time exceeding the sampling time $T_s$=1~$sec$. This clearly impedes the real-time implementation of the single-layer MPC. Moreover, vehicle speed profile can be accurately estimated using traffic and infrastructure information (V2I/V2X) only over a short horizon~\cite{sun2015integrating}. The prediction of the vehicle speed over an extended horizon is subject to uncertainties, which consequently affects the performance of the MPC for BTM over a long horizon. %The issue with computation complexity and information availability motivate us to seek the alternative control solution presented in Sections~\ref{sec:TrafficFlow} and \ref{sec:TwoMPC}.  %\vspace{-0.4cm} 
%
%\linespread{1.3}
% \begin{table}[h!]
% \begin{center}
% \small
% \centering
% \caption{Computation time comparison of the Single-Layer MPC with different prediction horizon for the results shown in Fig.~\ref{fig:UDDS_T_bat_SingleLayer_MPC_LongH_resize}. 
% \label{tab:Computation Time Comparison}} %\vspace{0.4cm}
% \begin{tabular}{lcccc}
%         \hline\hline
% ~~~~~~~~~~$N$& 10 & 60 & 120 & 180 \\ \hline
% $~~~average~CPU$      & 0.75 & 2.22 & 6.24 & 10.78 \\
% $time/iteration~[sec]$ & \textcolor{red}{-} & \textcolor{red}{$\uparrow$196\%} & \textcolor{red}{$\uparrow$732\%} &  \textcolor{red}{$\uparrow$1337\%}\\%\hline
% %
% % $~~~max~CPU$      &  &  &  &  \\
% % $time/iteration~[sec]$ & 2.40 & 4.51 & 9.57 & 18.47 \\
%         \hline\hline
%     \end{tabular}
% \end{center} \vspace{-0.5cm} 
% \end{table}
% \linespread{1}

\begin{figure}[h!]
 	\begin{center}
		\includegraphics[width=7.0 cm]{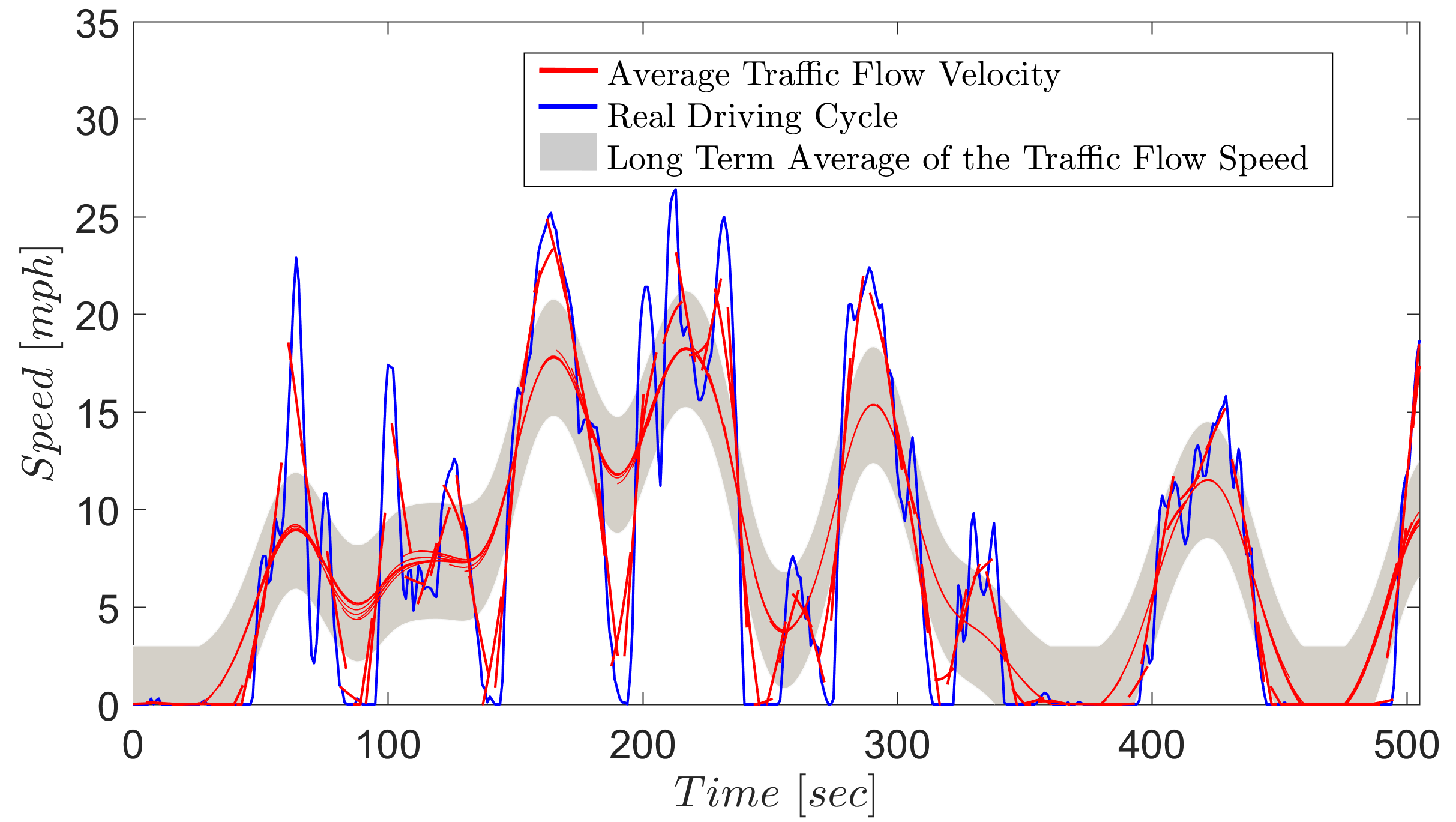} \vspace{-0.45cm}   % The printed column width is 8.4 cm.
 		\caption{The virtual traffic flow speed versus the actual driving cycle. In this paper, it is assumed that the traffic flow velocity information is repeatedly updated every cycle over a 250-$sec$ window. }\vspace{-0.6cm} 
 		\label{fig:NYCC_Estimate} 
 	\end{center}
\end{figure}

%\vspace{-0.4cm} 
%\section{Traffic Flow Model for Long Horizon Prediction} \label{sec:TrafficFlow}
%
While accurate long term vehicle speed prediction is difficult to obtain, it might not be necessary to claim most of the fuel saving benefits. We show in this paper that an approximate knowledge of future vehicle speed profile based on the average traffic flow velocity ($V_{flow}$) estimation can be integrated into the single-layer MPC controller to reduce energy consumption.~%While an exact demanded traction power over a short horizon is essential for the energy management of the electrified vehicles, it will be shown in this section that a minimum knowledge of the traffic flow information over a long horizon can be beneficial for BTM system optimization. Here, the demanded traction power of the vehicle over a long horizon is estimated with respect to the average traffic flow velocity ($V_{flow}$) according to the approach proposed in~\cite{sun2015integrating}. 
$V_{flow}$ is estimated according to the approach proposed in~\cite{sun2015integrating}, where the traffic flow data are extracted from a traffic monitoring system described in~\cite{herrera2010evaluation}, exploiting the extensive coverage of the cellular network, GPS-based position and velocity measurements, and the communication infrastructure of cellphones.%, see \cite{sun2015integrating} for more details on the actual vehicle speed versus the traffic flow speed. %In this paper, we assume that~%the traffic flow data are being collected, analyzed, and updated by a central/cloud server in real-time, and they are available to the vehicle control system at no extra computational cost. Additionally,
%$V_{flow}$ is updated every cycle as the traffic flow condition changes in the downstream. 
~Fig.~\ref{fig:NYCC_Estimate} illustrates the concept of the average traffic flow speed trajectory and compares it against the actual speed profile. It can be observed that the vehicle speed profile prediction is close to the actual speed for the first few cycles, before it merges into the average traffic flow speed (gray band) over the long horizon. For more details, see~\cite{sun2015integrating}. \vspace{-0.3cm} 
% %%
% \begin{figure*}
% 	\begin{center}
% 		\includegraphics[scale=0.4]{Results/UDDS_T_bat_LongH_SingleLayer_SecondLayer_180.eps} \vspace{-0.3cm}   % The printed column width is 8.4 cm.
% 		\caption{The performance of the Single-Layer and Two-Layer MPCs with accurate and estimated vehicle speed profiles over a long horizon for UDDS.}\vspace{-0.75cm} 
% 		\label{fig:UDDS_T_bat_LongH_SingleLayer_SecondLayer_180} 
% 	\end{center}
% \end{figure*}
%

% \newcommand{\greenline}{\raisebox{2pt}{\tikz{\draw[-,black!30!green,solid,line width = 0.9pt](0,0) -- (3mm,0);}}}
%
\newcommand{\greenline}{\raisebox{2pt}{\tikz{\draw[-,green,solid,line width = 0.9pt](0,0) -- (3mm,0);}}}
\newcommand{\blueline}{\raisebox{2pt}{\tikz{\draw[-,black!30!blue,solid,line width = 0.9pt](0,0) -- (3mm,0);}}}
\newcommand{\redline}{\raisebox{2pt}{\tikz{\draw[-,red,dashed,line width = 1.5pt](0,0) -- (4.5mm,0);}}}
\newcommand{\bluearrow}{\raisebox{2pt}{\tikz{\draw[->,blue,solid,line width = 1.5pt](0,0) -- (5mm,0);}}}
\newcommand{\rectangle}{\raisebox{0pt}{\tikz{\draw[black,solid,line width = 1.0pt](2.mm,0) rectangle (3.5mm,1.5mm);\draw[-,black,solid,line width = 1.0pt](0.,0.8mm) -- (5.5mm,0.8mm)}}}

%\vspace{-0.25cm} 
To understand the effects of uncertainties associated with the long prediction horizon, we consider the implementation of the single-layer MPC with both the actual speed profile, and with the one estimated from the traffic flow information. ~% , and the single-layer MPC results from Fig.~\ref{fig:UDDS_T_bat_SingleLayer_MPC_LongH_resize} with the prediction horizon $N=180~sec$.~%Despite the mismatch between the real vehicle speed profile and vehicle speed profile based on the traffic flow information%(Fig.~\ref{fig:NYCC_Estimate})
It was previously observed from Fig.~\ref{fig:UDDS_T_bat_SingleLayer_MPC_LongH_resize} that the single-layer MPC with exact vehicle peed profile results in up to $5\%$ saving in the battery energy compared to the rule-based controller. On the other hand, up to $3.9\%$ energy saving is achieved when using traffic flow information, compared to 5\% with the exact speed profile. Despite the mismatch between the real vehicle speed profile and long range prediction of the vehicle speed profile based on the traffic flow information, the energy saving is still substantial. Thus, the estimated long range vehicle speed profile via the traffic flow speed data can serve the purpose of the BTM system optimization.~%(Fig.~\ref{fig:NYCC_Estimate}), given the relatively slow dynamics of the thermal system, the average estimate of the traffic flow speed over the long horizon can serve the purpose of the BTM system optimization. On the other side, it can be seen that the results of $T_{bat}$ control is different. In both cases, the battery temperature is enforced to remain under the limit. However, 
%At the same time, the single-layer MPC using the exact vehicle speed profile results in less frequent constraint violation (by 74\%) and lower average battery temperature (by 4\%), compared to that using vehicle speed profile based on traffic flow information.%with the single-Layer MPC with average traffic floe seed data. Fig.~\ref{fig:UDDS_T_bat_LongH_SingleLayer_SecondLayer_180} shows the results of evaluating the single-layer MPC (Eq.~(\ref{eqn:Eq11})) with a long horizon ($N=180~sec$) and with the exact speed profile and with the speed profile extracted from traffic flow information,  together with results of the on/off rule-based controller as the baseline.
\vspace{-0.2cm} 

\section{Two-Layer MPC for Battery Thermal and Energy Optimization} \label{sec:TwoMPC} 
\begin{figure}[t!]
	\begin{center}
		\includegraphics[width=7.0 cm]{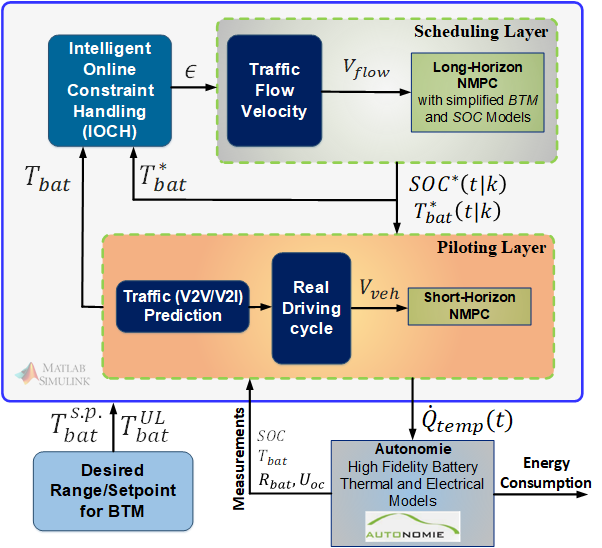} \vspace{-0.35cm} 
		\caption{Schematic of Two-Layer NMPC for battery thermal and energy management optimization.}\vspace{-0.95cm} 
		\label{fig:TwoLayer_MPC_Schematic} 
	\end{center}
\end{figure}

\vspace{-0.15cm} 
%Fig.~\ref{fig:UDDS_T_bat_LongH_SingleLayer_SecondLayer_180} shows the CPU computation time per iteration for
%It was shown that the traffic flow information over a long horizon can be used for battery thermal and energy management optimization. However, the computational burden is still considerable. 
It was observed that for the single-layer MPC, the maximum computation time per controller execution can reach as much as $20~sec$, even on a powerful 2.60 GHz processor (note that $T_s$=1~$sec$). In order to leverage the energy saving potentials of the long horizon traffic flow information and reduce the computation time of the MPC, we propose a two-layer MPC for battery thermal and energy optimization with a scheduling layer and a piloting layer as illustrated in Fig.~\ref{fig:TwoLayer_MPC_Schematic}. The scheduling layer has a relatively long horizon ($H_l$), and the dynamic of the system is sampled at a slower rate ($T_l>T_s$). The piloting layer has a short prediction horizon $H_s$, with a sampling time of $T_s$.\vspace{-0.15cm} 

\subsection{Scheduling Layer MPC with Long Horizon}\vspace{-0.1cm} 
In order to make the long-horizon MPC computationally efficient, the dynamic models for $\dot{T}_{bat}$ and $\dot{SOC}$ are simplified. To this end, first, $I_{bat}$ equation from (\ref{eqn:Eq4}) is approximated by using the Taylor series expansion with different accuracies for $SOC$ ($I_{bat}^{SOC}$), and the current ($I_{bat}^{T_{bat}}$)~\cite{bauer2014thermal}: 
\vspace{-0.15cm} 
%\small
\begin{gather}
\label{eqn:Eq12}
I_{bat}^{SOC}(t)=\frac{a_c\dot{Q}+P_{trac}}{U_{oc}}+\frac{R_{bat}(a_c\dot{Q}+P_{trac})^2}{U_{oc}^3},
\end{gather}
\vspace{-0.75cm} 
\begin{gather}
\label{eqn:Eq13}
I_{bat}^{T_{bat}}(t)=\frac{a_c\dot{Q}+P_{trac}}{U_{oc}}.
\end{gather}
\normalsize
%
%\doublespacing
%
Next, Eqs.~(\ref{eqn:Eq12}) and (\ref{eqn:Eq13}) are used to re-write the discretized battery state-of-charge ($f_{SOC,l}$) and temperature ($f_{T_{bat},l}$) dynamics over the long horizon $H_l$ assuming the sampling time $T_l$, where $SOC_l$ and $T_{bat,l}$ designate the states of the simplified model used in the scheduling layer.
%
% \vspace{-0.85cm} 
% \small
% \begin{gather}
% \label{eqn:Eq{14}}
% {SOC}_{l}(k+1)={SOC}_{l}(k)+~~~~~~~~~~~\\
% T_l\Big(\frac{R_{bat}(a_c\dot{Q}+P_{trac})^2+U_{oc}^2(a_c\dot{Q}+P_{trac})}{C_{nom}U_{oc}^3}\Big), \nonumber
% \end{gather}
% %
% \vspace{-0.5cm} 
% %
% \begin{gather}
% \label{eqn:Eq{15}}
% {T}_{bat,l}(k+1)={T}_{bat,l}(k)+~~~~~~~~~~~~~~~~~~~~~~~~~~~~~~\\
% T_l\Big(\frac{\dot{Q}(2a_cR_{bat}P_{trac}+U_{oc}^2)+R_{bat}P_{trac}^2}{C_{th, bat}U_{oc}^2}\Big) \nonumber
% \end{gather}
% \normalsize
%
%where $SOC_l$ and $T_{bat,l}$ designate the states of the simplified model used in the scheduling layer. 
The scheduling layer MPC is based on the following optimization problem formulation: \vspace{-0.45cm} 
%-------------------------------------
\begin{gather} \label{eqn:Eq16}
\begin{aligned}
& \underset{\dot{Q}(\cdot|k)}{\text{min}} & & \sum_{i=0}^{H_l} P_{temp}(i|k),  \\
%\tilde{\boldsymbol{y}}(k) & \\
& \text{s.t.} & & \text{constraints listed in (\ref{eqn:Eq11})}.
% & & T_{bat,l}(i+1|k)=f_{T_{bat},l}(i|k),~{ i=0,\cdots,H_l},\\
% & 
% & & SOC_l(i+1|k)=f_{SOC,l}(i|k),~{ i=0,\cdots,H_l},\\
% & 
% & &T_{bat}^{LL}\leq T_{bat,l}(i|k) \leq T_{bat}^{UL},~{ i=0,\cdots,H_l},\\
% & 
% & &30\% \leq SOC_l(i|k)\leq 90\%,~{ i=0,\cdots,H_l},\\
% % \end{aligned} 
% % \end{gather}
% % %
% % \begin{gather} \label{eqn:Eq16}
% % \begin{aligned}
% % & 
% % & &T_{bat}^{LL}\leq T_{bat,l}(i|k) \leq T_{bat}^{UL},~{ i=0,\cdots,H_l},\\
% % & 
% % & &30\% \leq SOC_l(i|k)\leq 90\%,~{ i=0,\cdots,H_l},\\
% & 
% & &-3000~W\leq\dot{Q}(i|k)\leq 0,~{ i=0,\cdots,H_l-1},\\
% & 
% & & T_{bat,l}(0|k)=T_{bat}(k),~SOC_l(0|k)=SOC(k).%\\
% %& 
%& & \dot{Q}^*(0|k)=\dot{Q}(k)   %\vspace{-0.5cm}
\end{aligned} 
\end{gather}
%--------------------------------------
The scheduling layer MPC optimizes $\dot{Q}$ over the long horizon, and its solution is used to schedule the desired values of the battery temperature ($T_{bat}^*$) and state-of-charge ($SOC^*$) for the piloting layer. Note that the structure of the scheduling layer MPC is similar to the single-layer MPC in (\ref{eqn:Eq11}), but with a different sampling rate. This, as will be shown later, results in significantly reduced computational load.%, as will be shown later. %However, as it will be shown later, the MPC of the scheduling layer is computationally faster than the single-layer MPC.%, and (iii) the demanded traction power profile, which is estimated with respect to the average traffic flow speed (fig.~\ref{fig:NYCC_Estimate}), has a significantly more smooth shape than the actual traction power profile used by the Single-Layer MPC; thus, the solver spends less time to find the solution of the optimization problem. 

\vspace{-0.2cm}
\subsection{Piloting Layer MPC with Short Horizon}\vspace{-0.1cm}
$T_{bat}^*$ and $SOC^*$ from the scheduling layer are passed on to the piloting layer, where these values are used by a short-horizon MPC for tracking. The output of the scheduling layer MPC is updated every $T_l$, during which the output of the short-horizon MPC is updated $\tau=T_l/T_s$ times, where $\tau\in \mathbb{Z}$ is the ratio between the long and short horizon length. The length of the scheduled values which need to be passed on to the piloting layer depends on the piloting layer prediction horizon ($H_s$). Moreover, since $T_l>T_s$, the scheduled $T_{bat}^*$ and $SOC^*$ are passed on as piece-wise constant functions: $T_{bat}^*(t|k)$ and $SOC^*(t|k)$~\cite{lefort2013hierarchical}. %This means that the reference trajectories passed to the piloting layer are assumed to be constant over a $\xi.T_s~sec$ period. 
%Fig.~\ref{fig:Task5_Integrated_HVAC_BTM} shows the communication process between the two layers, and how the scheduled references are passed on to the piloting layer from the scheduling layer. %\vspace{-0.5cm} 
%

The short-horizon MPC of the piloting layer is formulated as follows to track the scheduled $T_{bat}^*$ and $SOC^*$ references from the scheduling layer: \vspace{-0.20cm} 
%-------------------------------------
\begin{gather} %\label{eqn:Eq{17}}
\begin{aligned}
% & \underset{\dot{Q}(\cdot|k)}{\text{min}} & & \sum_{j=0}^{H_s} \left\lbrace\Big((T_{bat}(j|k)-T_{bat}^*(j|k)\Big)^2 \right.\\
% & 
% & &\left.~~~~~+w_1\big(SOC(j|k)-SOC^*(j|k)\Big)^2 \right\rbrace, \nonumber\\
& \underset{\dot{Q}(\cdot|k)}{\text{min}} & & \sum_{j=0}^{H_s} \Bigg\{ \begin{gathered} \big(T_{bat}(j|k)-T_{bat}^*(j|k)\big)^2\\+w_1\big(SOC(j|k)-SOC^*(j|k)\big)^2  \end{gathered}\Bigg\}, \nonumber\\
%\end{aligned}
%\end{gather}
%
%\begin{gather} \label{eqn:Eq{17}}
%\begin{aligned}
%\tilde{\boldsymbol{y}}(k) & \\
& \text{s.t.}
& & T_{bat}(j+1|k)=f_{T_{bat}}(j|k),~{ j=0,\cdots,H_s},\\
& 
& & SOC(j+1|k)=f_{SOC}(j|k),~{ j=0,\cdots,H_s},\\
%& 
%& &T_{bat}(i|k)\in\mathcal{T},~SOC(i|k)\in\mathcal{SOC},~{\scriptstyle i=0,\cdots,H_l}\\
& 
& &-3000~W\leq\dot{Q}(j|k)\leq 0,~{ j=0,\cdots,H_s-1},\\
& 
& & T_{bat}(0|k)=T_{bat}(k),~SOC(0|k)=SOC(k).%\\
%& 
%& & \dot{Q}(0|k)=\dot{Q}(k)
\end{aligned}
\end{gather}
%--------------------------------------
% \vspace{-0.35cm} 
% \begin{figure}[h!]
%  	\begin{center}
%  		\includegraphics[width=7.8 cm]{Results/TwoLayer_MPC_Block_Schematic_v2.eps} \vspace{-0.35cm}   % The printed column width is 8.4 cm.
%  		\caption{Two-layer hierarchical structure scheme. The calculated $T_{bat}^{*}$ and $SOC^{*}$ references are sent in a block form from the scheduling layer to the piloting layer, where the short-horizon NMPC tracks these references.}\vspace{-0.75cm} 
%  		\label{fig:Task5_Integrated_HVAC_BTM} 
% \end{center}
% \end{figure}

When performing optimization in the piloting layer, we assume that the vehicle speed and the demanded traction power can be estimated accurately over the short horizon. Moreover, the constraints on the battery temperature and $SOC$ are not considered, as the long-horizon MPC has enforced these constraints in the scheduling layer. This will help to (i) reduce the computation time of the piloting layer optimization problem, (ii) avoid the infeasibility problem of the short-horizon MPC. Moreover, the computation times of the two-layer MPC at both layers are significantly lower than the single-layer MPC, because of: 
\begin{itemize}
\item The larger sampling time used at the scheduling layer leading to substantially reduced optimization variables for the same prediction time window. %time length of the prediction horizon, the scheduling layer MPC deals with less optimization variables.
\item Simplified dynamics of $SOC_l$ and $T_{bat,l}$.
\item Non-redundant constraint enforcement with the battery temperature constraint being enforced at the scheduling layer and removed from the piloting layer MPC. %Since $T_{bat}$ is regulated to be very close to $T_{bat}^{UL}$ to save energy, the solver of the single-layer MPC spends more time to enforce $T_{bat}^{UL}$ constraint. On the other side, for the short-horizon MPC of the piloting layer, while the temperature is close to $T_{bat}^{UL}$, since there is no constraint on $T_{bat}$, the optimization problem just for tracking $SOC^*$ and $T_{bat}^*$ can be solved much faster.
\end{itemize}

%In this study, the long-horizon MPC of the scheduling layer has a prediction horizon of $H_l=50$ ($250~sec$) and, the short-horizon MPC runs with $H_s=10$ (10~sec). (\blueline) 
%{The energy saving result and computation times of the two-layer MPC are also compared with the single-layer MPC in Fig.~\ref{fig:UDDS_T_bat_LongH_SingleLayer_SecondLayer_180}. Using the same traffic flow information for both two-layer (\blueline) and single-layer (\redline) MPCs, the comparison results show that the energy saving potentials from both MPCs are the same $3.9\%$ improvement compared to the baseline.
\vspace{-0.35cm} 

\subsection{Intelligent Online Constraint Handling}\vspace{-0.1cm}

Since the battery temperature constraint is removed from the short-horizon MPC, an intelligent online constraint handling (IOCH) algorithm is added to the structure of the two-layer MPC (shown in Fig.~\ref{fig:TwoLayer_MPC_Schematic}) to reduce the battery temperature limit violation that may be caused by the mismatch between the actual driving conditions and assumed driving conditions based on average traffic flow information. The IOCH block takes into account the violation of the battery temperature limits, and monitors the difference between $T_{bat}$ and $T_{bat}^*$. The addition of the IOCH block introduces an extra optimization variable $\epsilon$ which modifies the upper temperature bound at the scheduling layer to avoid temperature limit violations at the piloting layer, with consideration of the battery cooling power minimization. For this purpose, the cost function of the long-horizon MPC at the scheduling layer in Eq.~(\ref{eqn:Eq11}) is modified as follows: \vspace{-0.25cm}
\begin{gather}
\label{eqn:Eq18}
\begin{aligned}
& \underset{\dot{Q}(i|k),\epsilon(k)}{\text{min}} & & \sum_{i=0}^{H_l} P_{temp}(i|k)+\gamma(\delta(T_{bat},T_{bat}^{UL})-\epsilon(i|k))^2,
\end{aligned}
\end{gather}
where, $\gamma$ is a weighting factor to adjust the controller effort for reducing the constraint violation. The long-horizon MPC is subject to the constraints listed in (\ref{eqn:Eq16}), except for $T_{bat}(i|k)$, which is now subject to: \vspace{-0.15cm}
\begin{gather}
\label{eqn:Eq19}
T_{bat}^{UL}(i|k)\leq T_{bat}^{UL}-\epsilon(i|k),~{ i=0,\cdots,H_l}\\
\epsilon(i|k)\geq 0,~{ i=0,\cdots,H_l-1}, \nonumber
\end{gather}
where, $T_{bat}^{UL}(i|k)$ denotes the variable upper limit of the battery temperature operation. The function $\delta(T_{bat},T_{bat}^*)$ in Eq.~(\ref{eqn:Eq18}) is defined as follows: 
\vspace{-0.2cm}
\begin{gather}
\label{eqn:Eq20}
\delta(T_{bat},T_{bat}^{UL})=\begin{cases}
0 & if~ T_{bat}\leq T_{bat}^{UL}\\
T_{bat}-T_{bat}^{UL}  & if~ T_{bat}>T_{bat}^{UL}
\end{cases}
\end{gather}
\begin{figure}[b!]
	\begin{center}
        \includegraphics[width=8 cm]{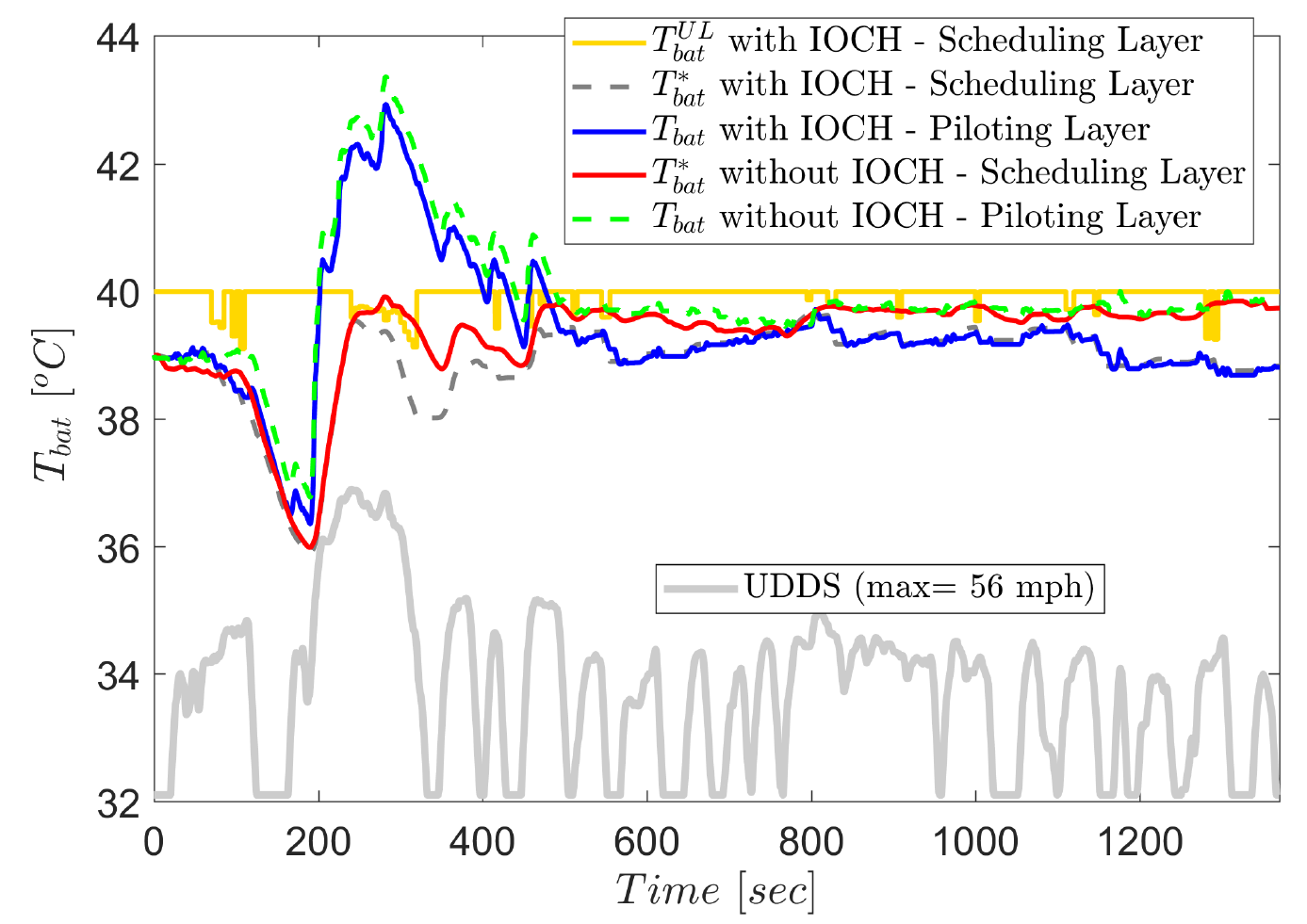} \vspace{-0.4cm} 
		\caption{The results of battery thermal management from Two-Layer MPC with and without IOCH for the UDDS ($T_{bat}^{UL}$=$40^oC$,~$T_{bat,0}$=$39^oC$).}\vspace{-0.2cm} 
		\label{fig:UDDS_T_bat_TwoLayer_ini39_V2} 
	\end{center}
\end{figure}

Fig.~%\ref{fig:UDSS_T_bat_TwoLayer} and 
\ref{fig:UDDS_T_bat_TwoLayer_ini39_V2} shows the performance of the two-layer MPC for UDDS at $T_{bat,0}=39^oC$, $H_l=180~sec$, and $H_s=15~sec$. It can be seen that when $T_{bat}^{UL}$ is constant and set to be $40^oC$, the scheduled battery temperature trajectory ($T_{bat}^*$) does not violate the upper limit constraint. But, the actual battery temperature does because of the mismatch between the real and predicted driving cycle. However, by using Eq.~(\ref{eqn:Eq18}) as the cost function of the scheduling layer MPC, the added optimization variable $\epsilon$ modifies the upper limit (Eq.~(\ref{eqn:Eq19})) with respect to the mismatches between $V_{flow}$ and $V_{veh}$, and reduces the overall $T_{bat}^{UL}$ violation by $13\%$. It should be noted from Fig.~\ref{fig:UDDS_T_bat_TwoLayer_ini39_V2} that compared to the two-layer MPC without IOCH, the energy saving results (as determined by terminal battery SOC) of the two-layer MPC with IOCH is $1\%$ lower, which is expected, as the overall battery temperature is lower and constraint violations are less frequent. This is the price paid for putting more effort to maintain $T_{bat}$ within the desired range. While temporary violations of $T_{bat}^{UL}$ are tolerated, it is required that $T_{bat}$ to be maintained withing the optimum operating range to improve the battery life and health in long term~\cite{neubauer2014thru}.\vspace{-0.15cm}%Sec.~\ref{sec:4}. %As discussed before, these mismatches often occur during sharp accelerations and deceleration events, where the average traffic flow speed information does not reflect those aggressive moments. 
\vspace{-0.1cm} 
%

%
%The effectiveness of the proposed constraint handling technique, and its impact on the Two-Layer MPC computation time will be further discussed in Section~\ref{sec:4}.
%\vspace{-0.3cm}
\section{Two-Layer MPC Simulation Results }\label{sec:4}
\begin{figure}[h!]
	\begin{center}
\includegraphics[width=7.5 cm]{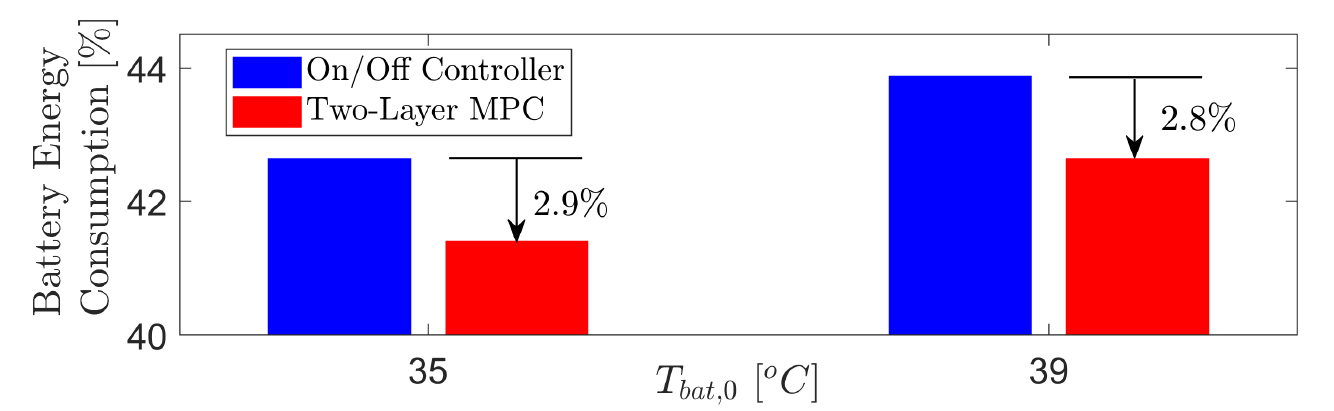}\vspace{-0.4cm}   % The printed column width is 8.4 cm.
		\caption{The results of battery energy management by using Two-Layer MPC with IOCH and On/Off controller ($T_{bat}^{s.p.}$=$35^oC$) for UDDS at $T_{bat,0}$=$35^oC$ and $39^oC$.}\vspace{-0.8cm} 
		\label{fig:UDDS_SOC_SingleLayer_Tracking_Comparison} 
	\end{center}
\end{figure}

\begin{figure}[h!]
	\begin{center}
		\includegraphics[width=7.5 cm]{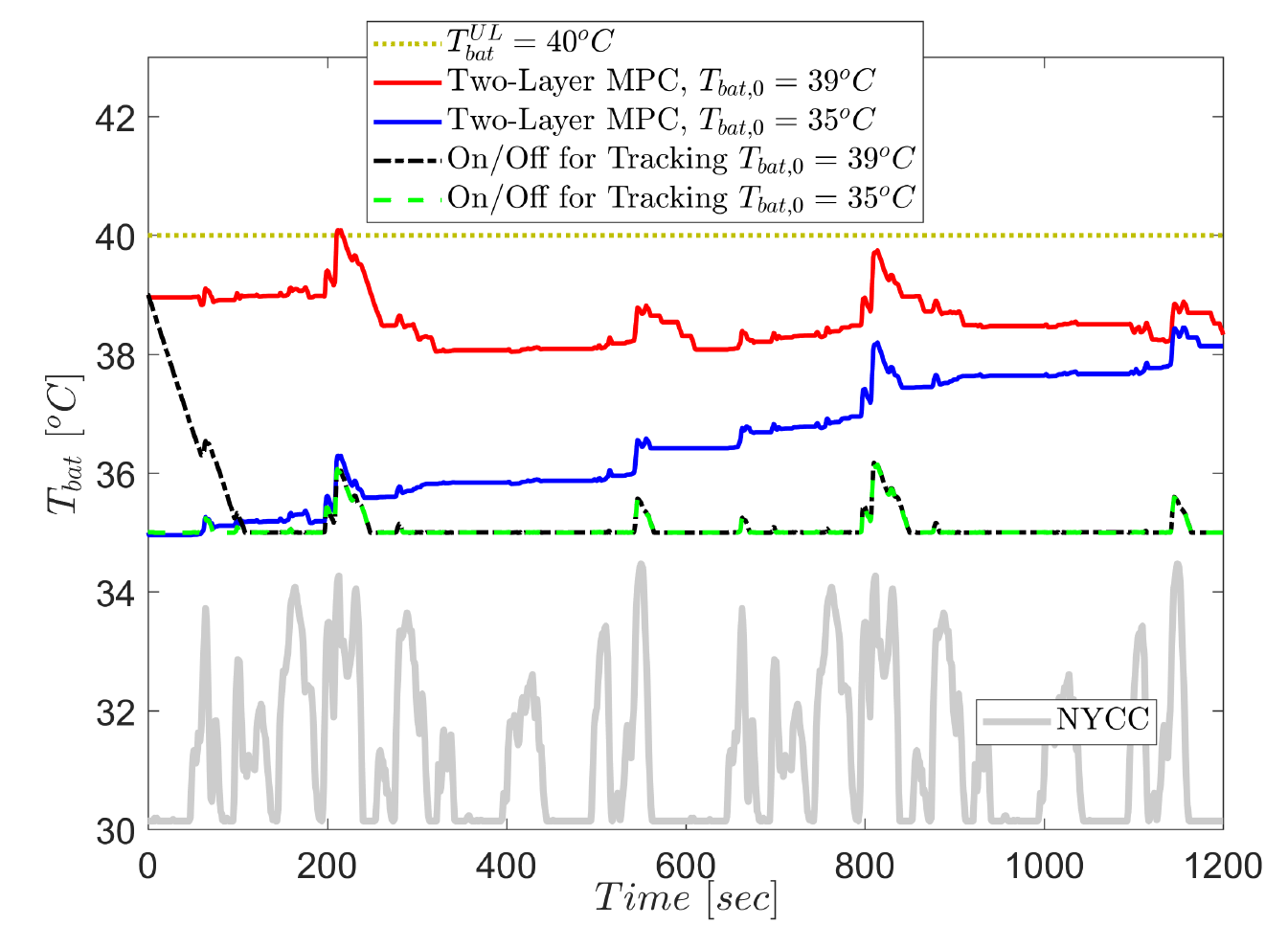} \vspace{-0.4cm}   % The printed column width is 8.4 cm.
		\caption{The results of BTM by using Two-Layer MPC with IOCH and On/Off controller ($T_{bat}^{s.p.}$=$35^oC$) for NYCC at $T_{bat,0}$=$35^oC$ and $39^oC$.}\vspace{-0.55cm} 
		\label{fig:NYYC_T_bat_SingleLayer_Tracking_Comparison} 
	\end{center}
\end{figure}

\begin{figure}[h!]
	\begin{center}
\includegraphics[width=7.7 cm]{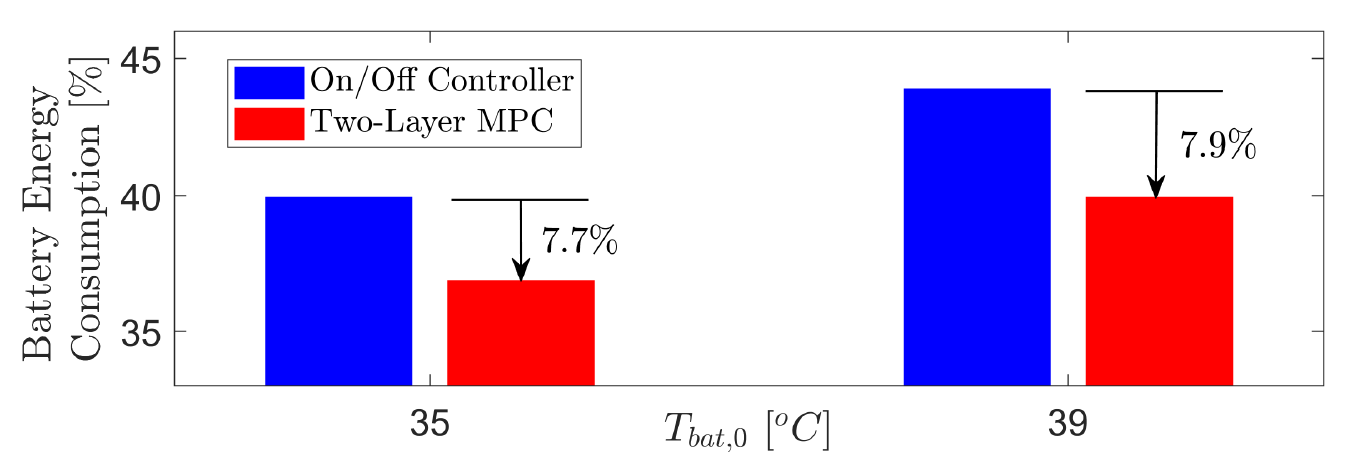} \vspace{-0.4cm}   % The printed column width is 8.4 cm.
		\caption{The results of battery energy management by using Two-Layer MPC with IOCH and On/Off controller ($T_{bat}^{s.p.}$=$35^oC$) for NYCC at $T_{bat,0}$=$35^oC$ and $39^oC$.}\vspace{-0.75cm} 
		\label{fig:NYYC_SOC_SingleLayer_Tracking_Comparison} 
	\end{center}
\end{figure}

In order to demonstrate the energy saving potentials of the proposed two-layer MPC for battery thermal management of electric vehicles, the predictive controller with IOCH is simulated for UDDS and the New York City Cycle (NYCC) at different initial battery temperature conditions. The NYCC features low speed stop-and-go traffic conditions. These results are compared with the traditional rule-based controller, and presented in Figs.~\ref{fig:UDDS_SOC_SingleLayer_Tracking_Comparison}-\ref{fig:NYYC_SOC_SingleLayer_Tracking_Comparison}. For UDDS, it can be seen from Fig.~\ref{fig:UDDS_SOC_SingleLayer_Tracking_Comparison} that by the end of the driving cycle, the two-layer MPC reduces the drop in the battery $SOC$ by {$2.9\%$} and $2.8\%$ for $T_{bat,0}=35^oC$ and for $T_{bat,0}=39^oC$, respectively, compared to the rule-based controller. %Figs.~\ref{fig:UDDS_T_bat_SingleLayer_Tracking_Comparison} shows that despite $T_{bat}^{UL}$ violations during the period when demanding traction power occurs, the two-layer MPC brings back the temperature to below the target limit, and keeps the temperature close to this limit to minimize the required cooling power from the battery.~%

%
% \begin{figure}[h!] %\vspace{-0.5cm}
% 	\begin{center}
% 		\includegraphics[width=8 cm]{Results/UDDS_T_bat_SingleLayer_MPC_Modified_Final_resize_v2.eps} \vspace{-0.4cm}   % The printed column width is 8.4 cm.
% 		\caption{The results of BTM by using Two-Layer MPC with IOCH and On/Off controller ($T_{bat}^{s.p.}$=$35^oC$) for UDDS at $T_{bat,0}$=$35^oC$ and $39^oC$.}\vspace{-0.5cm} 
% 		\label{fig:UDDS_T_bat_SingleLayer_Tracking_Comparison} 
% 	\end{center}
% \end{figure}
%UDDS_SOC_SingleLayer_MPC_Modified_Final_resize_v3_addon
\vspace{-0.3cm} 
\vspace{-0.2cm} 

For the NYCC, since the vehicle speed is low on average, aggressive rises in the demanded traction power and battery temperature are not observed. Thus, the two-layer MPC is able to maintain the battery temperature well below the upper limit, even with $T_{bat,0}=39^oC$. By looking into the battery energy management results for the NYCC in Fig.~\ref{fig:NYYC_SOC_SingleLayer_Tracking_Comparison}, it can be seen that the conservative design of the rule-based controller results in a significant drop in the battery $SOC$ by the end of the driving cycle. Compared to the rule-based controller, the two-layer MPC reduces battery energy consumption (as determined by terminal $SOC$) by {$7.9\%$} and $7.7\%$ for $T_{bat,0}=39^oC$ and $T_{bat,0}=35^oC$, respectively. This is because the rule-based controller tries to maintain the temperature at $T_{bat}^{s.p.}=35^oC$, regardless of the traffic and driving conditions. %While the maximum demanded traction power during NYYC is lower than that of UDDS, the rule-based BTM demands for a relatively high tempering power. 
%Unlike the rule-based controller, the two-layer MPC, which anticipates future vehicle speed profile based on the long-term traffic flow information, maintains $T_{bat}$ within the desired operation range, while minimizing the cooling power. 
%\vspace{-0.35cm}

Finally, Table~\ref{tab:Computation_Time_Comparison_2} summarizes the average and maximum required computation times for each layer of the two-layer MPC with and without IOCH. Also the average computation time of the single-layer MPC from Sec.~\ref{sec:3} with a similar prediction horizon as of the scheduling layer MPC ($N=H_l$), and $T_s$=1~$sec$ is listed in Table~\ref{tab:Computation_Time_Comparison_2}.~%It can be seen that the addition of the extra optimization variable $\epsilon$ to the cost function of the scheduling layer MPC increases the computation time by 142\%. Despite this increase, 
The average and maximum computation times of the scheduling layer MPC with and without IOCH are less than the sampling time ($T_l$=$5~sec$). %Since the Two-Layer MPC with variable $T_{bat}^{UL}$ is keeping the battery temperature further away from the target upper limit,
%Table~\ref{tab:Computation_Time_Comparison_2} also shows that the average computation time of the piloting layer MPC with IOCH is slightly reduced by $4\%$, compared to the same controller without IOCH. 
Moreover, it can be seen that in all cases, the average and maximum computation times of the piloting layer MPC is less than the sampling time $T_s$=$1~sec$. Overall, it can be concluded that unlike the single-layer MPC, the two-layer MPC provides a computationally efficient framework.%, which allows for utilization of the long-term traffic information in the body of the electric vehicle battery thermal and energy management system in real-time. 
\vspace{-0.15cm} 
\begin{table}[h!]
%\linespread{0.8}
\begin{center}
\small
\centering 
\caption{Computation time comparison of the Single-Layer and Two-Layer MPCs. 
\label{tab:Computation_Time_Comparison_2}} %\vspace{0.4cm}
\begin{tabular}{lcc}
\textbf{Controller} & {$T_{bat}^{UL}=40^oC$} & {$T_{bat}^{UL}=variable$}  \\ 
        \hline\hline
 \textcolor{red}{Single Layer MPC} & \textcolor{red}{$N=180~sec$} & \textcolor{red}{$T_s=1~sec$}  \\ %\hline
$Average~CPU~Time$      & 10.78 (sec) & N/A  \\
   \hline\hline
\textcolor{blue}{Scheduling Layer MPC} & \textcolor{blue}{$H_l=180~sec$} & \textcolor{blue}{$T_l=5~sec$}  \\
$Average~CPU~Time$      & 0.855 (sec) & 2.071 (sec)  \\
%$Time/Iteration~[sec]$ & \textcolor{red}{-} & \textcolor{red}{$\uparrow$142\%} \\
%\hline
%
$Max~CPU~Time$      & 1.396 (sec) &4.265 (sec)    \\
%$time/iteration~[sec]$ & 1.396 &4.265 \\
        \hline\hline
\textcolor{blue}{Piloting Layer MPC} & \textcolor{blue}{$H_s=15~sec$} & \textcolor{blue}{$T_s=1~sec$} \\ %\hline
$Average~CPU~Time$      & 0.206 (sec) & 0.197 (sec)  \\
%$Time/Iteration~[sec]$ & \textcolor{green}{-} & \textcolor{black!30!green}{$\downarrow$-4\%} \\
%\hline
%
$Max~CPU~Time$   &  0.563 (sec) & 0.780 (sec)   \\
%$time/iteration~[sec]$ & 0.563 & 0.780 \\
        \hline\hline
    \end{tabular}
\end{center} \vspace{-0.55cm} 
\end{table}
\linespread{1}
%
%\vspace{-0.45cm}
% \begin{figure}[h!]
% 	\begin{center}
% 		\includegraphics[width=\columnwidth]{Results/UDDS_MPC_Tracking_Tbatdes30.eps} \vspace{-0.7cm}   % The printed column width is 8.4 cm.
% 		\caption{The results of battery thermal management by using the two-layer NMPC for tracking, and simple MPC for tracking for UDDS at $T_{bat,0}=35^oC$ for $T_{bat}^{s.p.}=30^oC$. }\vspace{-0.3cm} 
% 		\label{fig:UDDS_MPC_Tracking_Tbatdes30} 
% 	\end{center}
% \end{figure}

%\subsection{Proposed A/C System Prediction Model and Verification}

\section{Summary and Conclusions}\label{sec:5}
This paper investigates the design of a predictive and integrated battery thermal management (BMT) system in a connected and automated vehicles environment to improve energy efficiency and extend range of battery electric vehicles. To this end, first a single-layer MPC formulation was proposed to minimize the battery cooling power, while enforcing the battery temperature to be within the desired range. The simulation results showed that due to the relatively slow thermal dynamics of the battery, the MPC for BTM system requires information over a long horizon to achieve the design objectives. The simulation results confirmed that inclusion of the long horizon vehicle speed profile leads to $3-8\%$ saving in the total battery energy consumption, specifically for congested city driving cycles, e.g., NYCC. However, the single-layer MPC with a long horizon is computationally demanding. A two-layer MPC with an intelligent online constraint handling (IOCH) was then developed to (i) reduce the computation time as compared to the single-layer MPC, (ii) utilize the long-term traffic flow information along with the short-term vehicle speed predictions, and (iii) account for the mismatch between the actual vehicle speed profile and the predicted traffic flow speed used over the long horizon to reduce the overall battery temperature limit violation. The simulation results showed that the proposed two-layer MPC is able to achieve energy consumption reduction at a lower computational cost and without relying on the precise knowledge of the future vehicle speed profile. %By taking pro-active actions, the two-layer MPC reduces the battery temperature limit violations by $10-20\%$, compared to a rule-based controller. 
{Moreover, while up to $2.9\%$ energy saving was achieved for the UDDS, it was shown that for low-speed congested driving cycles, e.g., NYCC, higher energy saving can be achieved, and the two-layer MPC can save up to $7.9\%$ of the battery energy.}  
\vspace{-0.3cm}

% \section*{Acknowledgment}
% %\fi
% This material is based upon the work supported by the United States Department of Energy (DOE), ARPA-E NEXTCAR program under award No. DE-AR0000797. %Dr. Xun Gong from University of Michigan, Ann Arbor is gratefully acknowledged for his technical comments during the course of this study. 
% \vspace{-0.155cm}
%%%%%%%%%%%%%%%%%%%%%%%%%%%%%%%%%%%%%%%%%%%%%%%%%%%%%%%%%%%%%%%%%%%%%%%%%%%%%%%%

\bibliographystyle{unsrt} % unsrt
\bibliography{ACC2018Ref.bib} \vspace{-0.25cm}

% \begin{appendices}
% \small
% \section{Model Parameters}
% \label{FirstAppendix}
% %\begin{table}
% \begin{scriptsize}
% \centering
% \label{my-label}
% \begin{tabular}{ll}
% $battery~modules=6~[-]$ & $battery~cells = 168~[-]$ \\
% $cell~mass=0.34~[kg]$ & $C_{nom} = 7200-10800~[Wh]$ \\
% $m_{bat}=2.0400~[kg/module]$ & $C_{th,bat} = 521~[J/kgK]$ \\
% % $m=1450~[kg]$ & $A_f=2.52~[m^2]$ \\
% % $C_d=0.28~[-]$ & $\rho=1.2~[kg/m^3]$ \\
% % $C_f=0.015~[-]$ & $a_c=33.34~[-]$ \\

% \end{tabular}
% %\end{table}
% \end{scriptsize}
% \end{appendices}

\end{document}